# Collision-Induced Dissociation Studies on Fe(O$_2$)$_n^+$ (n=1-6) Clusters: Application of a New Technique Based on Crossed Molecular Beams

*Michalis Velegrakis[*], Claudia Mihesan,[§] and Maria Jadraque*

Institute of Electronic Structure and Laser, Foundation for Research and Technology - Hellas, Heraklion 71110, Crete, Greece

[§] "Petru Poni" Institute of Macromolecular Chemistry, Centre of Advanced Research in Nanobioconjugates and Biopolymers, 700487 Iasi, Romania

**Abstract**

Gas-phase oxygen-rich iron oxide clusters Fe(O$_2$)$_n^+$ (n=1-6), are produced in a molecular beam apparatus. Their stability and structure are investigated by measuring the fragmentation cross sections from collision-induced-dissociation experiments. For this purpose, two different techniques have been employed. The first one relies on the measurement of the fragments resulting after collisional activation and subsequent dissociation of mass selected cluster ions in a molecular beam passing through a cell filled with noble gas atoms. The second one is a new approach that we introduce and is based on crossed molecular beams to measure the fragmentation cross sections, in a more efficient manner without mass selection of the individual clusters. The cross sections obtained with the different techniques are compared with each other as well as with theoretical ones resulting from the application of a simple geometrical projection model. Finally, the general trends observed are compared with results for other Fe-molecule clusters available in the literature.

[*]Corresponding author: vele@iesl.forth.gr
e-mails: mihesan@iesl.forth.gr, mjadraque@iqfr.csic.es

# 1. Introduction

The importance of iron oxide $Fe_xO_x$ clusters in various scientific and technological fields such as heterogeneous catalysis, surface chemistry, solid state physics, corrosion, biochemical oxygen transport, oxide film formation, magnetic nanomaterials etc. has triggered a wide range of experimental[1,2,3,4,5,6,7,8,9,10,11,12,13,14] and theoretical[4,13,15,16,17,18,19,20,21] studies regarding their formation, stability, structure or chemical and physical activity performed in different laboratories.

For the experimental studies the preferred method for producing iron oxide clusters is the laser ablation of an iron target in the presence of $O_2$ seeded in a carrier gas, followed by a supersonic expansion that cools the formed clusters [1-12]. The mixing of the laser ablated species with the $O_2$ in a closed space (cluster growth channel) forms, through multiple collisions, large $Fe_xO_y$ clusters with a broad range of stoichiometries that differ from the known bulk iron oxides (FeO, $Fe_2O_3$, $Fe_3O_4$), corresponding to the common oxidation states of Fe. These $Fe_xO_y$ clusters are usually described as oxygen-poor (x>y) or oxygen-rich (y>x) iron oxides and the number of oxygen atoms in $Fe_xO_y$ clusters is influenced by the content of $O_2$ in the carrier gas: a higher $O_2$ concentration leads to the formation of higher oxidized species. For instance, slightly O-rich clusters ($Fe_xO_x$, $Fe_xO_{x+1}$, and $Fe_xO_{x+2}$) have been obtained by Bernstein et al.[10,11] and their formation is favored by a higher $O_2$ concentration. The same trend was observed for anionic clusters: the number of oxygen atoms in $Fe_xO_y^-$ can be slightly varied by increasing the oxygen content of the carrier gas.[2]

In order to gain more information about the stability and the structure of the clusters one typical method is to supply them with an excess of energy leading to their fragmentation and then to identify/analyze the fragmentation products. Usually, the energy deposition in the cluster occurs either by photon absorption in photofragmentation experiments or by collisional activation with a gas in collision-induced dissociation (CID) experiments. The decomposition of $Fe_xO_y^+$ clusters (1≤x≤17,

y=x±1, 2, 3) after photoexcitation with laser light has been reported by Duncan et al.[5] Their results show the loss of excess oxygen, followed by a sequential elimination of FeO units. The fragmentation pathways of small $Fe_xO_y^+$ (x = 1–4, y ≤ 6) clusters, obtained by chemical ionization of $Fe(CO)_5/O_2$ mixtures, were studied by Schröder et al.[13] It was found that the loss of molecular oxygen upon collisional activation becomes important as the formal oxidation state (determined by the O/Fe ratio) of Fe is increasing. Reilly et al.[4] have investigated the fragmentation channels of collisionally activated $Fe_xO_y^+$ clusters with Xe atoms in a guided ion beam spectrometer. They showed that the O-rich series $FeO_y^+$ (y=1-10) lost successive $O_2$ units uncovering a $FeO^+$ core. CID experiments on iron oxide clusters $Fe_xO_y^+$ (x=1–3, y=1–6) with Xe were reported by Armentrout et al.[8] and correlations of the fragmentation patterns with the structure of the clusters have been made. For example, $FeO_4^+$ was found to easily dissociate to $FeO_2^+$ indicating the presence of a loosely bound $O_2$ molecule.

In a recent experimental and theoretical study[22] on iron oxide clusters, we reported on the formation of the oxygen-rich mono-iron $FeO_y^+$ clusters with y=1- 16, in a pure $O_2$ gas expansion over the plasma arising from the ablation of a Fe-target. Based on mass spectra and ab initio calculations, we showed that the clusters with an even number of oxygen atoms $Fe(O_2)_n^+$ (n=1-6) show increased stability and that $Fe(O_2)_5^+$ has prominent structural, thermodynamic and magnetic properties. These findings were further supported by preliminary results from collision-induced dissociation experiments.

In this work, we present a more thorough study of the stability and structure of $FeO_y^+$ clusters. Specifically, we applied the method of collision-induced dissociation to measure fragmentation cross sections employing two different approaches. The first one is the standard CID technique, where mass selected clusters collide with noble gas atoms in a collision chamber. The initially selected ion and all its possible fragments are separated and recorded with a reflectron time of flight mass spectrometer.

This CID of mass selected clusters is a well established technique (see Ref.s 23,24,25,26) and it was applied to gas-phase titanium oxide cluster in our laboratory.[27,28,29] In this study we introduce an alternative, faster and more efficient technique for measuring the CID processes in cluster atom collisions. Using crossed molecular beams with Ne as the collision partner, the fragmentation cross section and consequently the cluster stability are measured for the first time with this new technique. The technique is based on the rejection of all low kinetic energy fragments produced for each cluster after collisions with a noble gas atom.

## 2. Experimental apparatus and methods

### 2.1 Setup

The cluster source and the basic experimental setup (see Fig. 1a) have been described in detail elsewhere.[30,31,32] Here only the general features and some new modifications will be presented.

The iron-oxygen molecular aggregates are formed by mixing the laser-vaporized metal with a supersonic expansion of $O_2$ in vacuum. A Nd:YAG laser beam ( wavelength 1064 nm, pulse width 10 ns, repetition rate 10Hz) is focused onto a rotating pure Fe target and the ablation plasma plume is crossed perpendicularly a few millimeters above the target surface with the pulsed $O_2$ molecular jet produced by a homemade nozzle (diameter 0.5 mm, backing pressure 4 bar). Association reactions in the mixing region lead to the formation of oxygen-containing molecular aggregates.

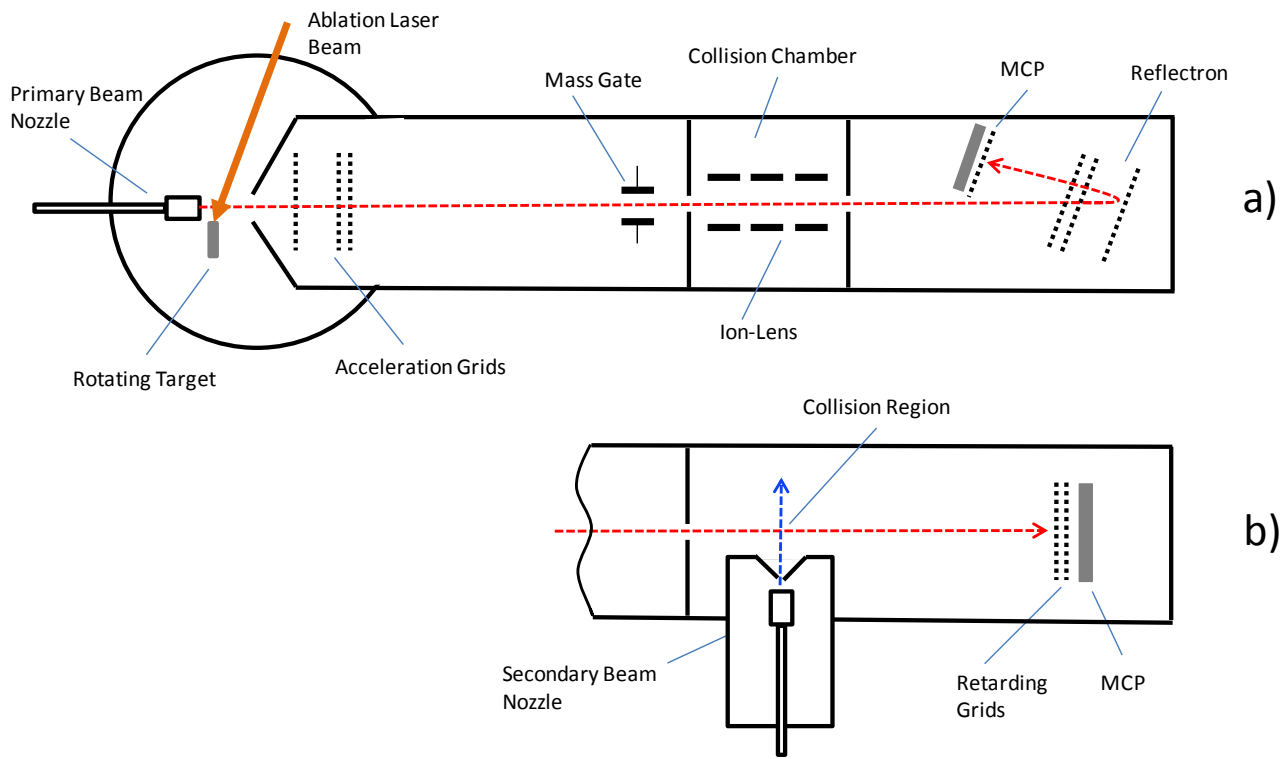

FIG. 1. Schematic diagram of the experimental arrangement indicating the two different configurations used to measure fragmentation cross sections by: a) "conventional" CID method with mass selection and reflectron detector and b) crossed molecular beams arrangement with a retarding potential analyzer.

Species already present or newly formed in the expansion are cooled and form a molecular beam that enters through a 4 mm diameter skimmer in the acceleration area of a time of flight (TOF) mass spectrometer. The positively charged species in the clusters beam are accelerated by a fast-switched potential pulse ($V_{acc}$=1500 Volt) at a laboratory kinetic energy $E_{LAB} = q.e.V_{acc}$ eV (e is the electron charge and q the charge state of a given ion) into the free flight zone of the time of flight mass spectrometer. The ions are detected with a multichannel plate (MCP) detector either after reflection from a reflectron assembly (Fig. 1a) or directly in a linear arrangement (Fig. 1b). The MCP signal is recorded by a 150MHz oscilloscope (LeCroy 9410) and the time of flight spectra are transferred through a GPIB interface to a PC. Software developed in our laboratory is used for data acquisition and processing.

### 2.2 Fragmentation processes

For studies on the stability and structure of the clusters we use the method of collisional activation of cluster ions with noble gas neutral atoms. In two-body formalism the main channels for the outcome of a cluster ion ($AB^+$) collision with a neutral atom (X) are:

$$AB^+ + X \rightarrow AB^{+(*)} + X^{(*)} \quad \text{(1) elastic and/or inelastic scattering}$$

$$\rightarrow A^+ + B + X \text{ or } A + B^+ + X \quad \text{(2) fragmentation}$$

$$\rightarrow AB + X^+ \quad \text{(3) charge transfer}$$

$$\rightarrow AB^{z+} + X + ze^- \quad \text{(4) multiionization (charge stripping)}$$

$$\rightarrow (AB^{z+})^* \rightarrow A^+ + B^{+(z-1)} \quad \text{(4a) Coulomb explosion}$$

The first reaction (1) indicates elastic or inelastic scattering (the asterisk denotes excited species), while the other reactions (2-4a) imply the formation of charged and neutral species possessing either masses lower than those of the initial ion or charge state z >1.
Neglecting the energy released from the dissociation (a few eV's), in cases, where a parent cluster with mass $m_P$ dissociates to a fragment with mass $m_F$, the laboratory kinetic energy $E_F$ of a fragment is lower than that of the parent $E_{LAB}$:

$E_F = (E_{LAB}/z) \cdot m_F/m_P$ (6).

This fact allows the separation of the parents and fragments by using devices sensitive to the energy of the incident ions. Such devices are the reflectron and the retarding potential analyzer used in this work in order to measure the fragmentation cross sections of the clusters by CID experiments, by two different experimental arrangements: beam-gas cell (Fig. 1a) and crossed molecular beams (Fig. 1b).

### 2.2.1 Beam-gas cell configuration

In the first case (see Fig. 1a), we apply the widely used procedure concerning fragmentation via collisions of several cluster systems.[23,24,25,26] We employed this method in small $Ti_xO_y^+$ clusters and the experimental details and results are described in a previous paper.[29] Briefly, mass selected cluster ions pass through a collision chamber filled with Kr gas at a pressure of $6 \times 10^{-4}$ mbar. Following dissociation, fragments with lower kinetic energy than the parent are recorded as different peaks in the reflectron TOF spectrum before the parent ion peak.

Under single collision conditions, the fragmentation cross section Q is determined from the parent intensity $I_P$, corresponding to parent peak in the TOF spectrum and the intensity $\Sigma I_F$ of all the fragment peaks using the relation:

$$Q = -\frac{1}{NL} \ln\left(\frac{I_P}{I_P + \sum I_F}\right)$$ (7),

where N is the number density of the target Kr-gas and L is the length of the collision cell. These parameters (N and L) cannot be accurately measured[33] in the present experimental arrangement but are common for all the clusters investigated and therefore are not necessary for the comparison of different clusters measured under the same conditions.

This well established CID method - called here "conventional" - is relatively simple but is time-consuming when measuring a series of clusters, due to the necessity of mass selection. It requires long term stability of the cluster beam and also special care has to be paid to avoid signal losses because of the different focusing conditions for each particular cluster, different fragments' flight paths through the reflectron, etc.

### 2.2.2 Crossed molecular beams

In this case the collision chamber (Fig. 1a) is removed and the cluster beam crosses perpendicularly a secondary molecular beam in the field-free zone of the mass spectrometer. The experimental arrangement is shown in Fig. 1b. The secondary beam contains Ne that expands from a nozzle similar to the primary beam nozzle (0.2 mm diameter, 3.5 bar backing pressure) and is placed in a separated chamber pumped with a 360 l/sec turbo pump, keeping the background pressure at ~$10^{-4}$ mbar during the operation. The secondary beam chamber is connected to the main TOF-tube with a 1 mm skimmer so that the two molecular beams cross perpendicularly in the scattering chamber that is pumped with two turbo pumps of 1000 l/sec and 150 l/sec in parallel and this allows for a background pressure of ~$10^{-6}$ mbar under operating conditions. The volume of the interaction region of the two beams is ~$1 \times 1 \times 1$ mm$^3$. The ion detector consists of a MCP with additionally a set of two grids in front of it. The first grid is grounded, while on the second one (closer to MCP) is applied a variable voltage V$_{grid}$. This voltage rejects any ion with a kinetic energy lower than $z \cdot e \cdot$V$_{grid}$ thus forming, a *retarding energy analyzer*. This device has an energy resolution better than 8%, which is the measured FWHM of the kinetic energy distribution of the main cluster ions formed.

In this way, the contribution of the different reaction channels (1-4a) can be measured. Specifically, a grid potential slightly lower (a few tens of volts) than the acceleration potential ($V_{grid} \lesssim V_{acc}$), will exclude signals arising from all the ionic fragments. Furthermore, the signal for $V_{grid}$ significantly higher than the accelerating voltage ($V_{grid} \gg V_{acc}$), corresponds to the amount of all neutrals produced in reactions (2-4), because *all charged species* are rejected and only neutral species can reach the detector. The impact of neutral molecules, moving with high velocity, with the anode of a MCP produces secondary electrons and thus the neutrals can be detected with the MCP.[34]

With these considerations, similar to light attenuation by an absorbing medium, the parent cluster's abundance before and after the interaction with the secondary beam can be described by Beer's law: $I = I_0 \exp(-QNL)$, where I is the (remaining) intensity of nonfragmented clusters, $I_0$ is their initial intensity, N is the secondary beam gas density, and L is the effective interaction length (width of the secondary beam). To account for fragmentation due to collisions of the clusters with background molecules and for possible metastable decay, the signals (integrated peak for each cluster in the TOF spectrum) with (ON) and without (OFF) the secondary beam have to be measured. Thus, $I_0 = I_{OFF}(V_{grid} \lesssim V_{acc}) - I_{OFF}(V_{grid} \gg V_{acc})$ and $I = I_{ON}(V_{grid} \lesssim V_{acc}) - I_{ON}(V_{grid} \gg V_{acc})$. The fragmentation cross section can then be evaluated from the relation:

$$Q = -\frac{1}{NL}\ln\left(\frac{I}{I_0}\right) = -\frac{1}{NL}\ln\left(\frac{I_{ON}(V_{grid} \lesssim V_{acc}) - I_{ON}(V_{grid} \gg V_{acc})}{I_{OFF}(V_{grid} \lesssim V_{acc}) - I_{OFF}(V_{grid} \gg V_{acc})}\right) \qquad (8).$$

Hence, to determine the fragmentation cross section, TOF spectra have to be measured at two different retarding potential settings: at $V_{grid} \lesssim V_{acc}$ and at $V_{grid} \gg V_{acc}$. Similar to Sec. 2.2.1, the parameters N and L cannot be measured but are considered the same for all the clusters due to an easily achieved good overlapping of the ~600 μs secondary beam pulse with the <60 μs interval needed for the primary beam to pass through the crossing region. Thus, it is straightforward to attain a very good temporal overlapping between the primary and secondary beam.

The timing of the experiment in both configurations, namely the synchronization of primary nozzle, ablation laser, accelerating voltage pulse and secondary nozzle (in crossed beams arrangement) is controlled by a digital delay/pulse generator (Stanford Research Systems Inc.) and is optimized for maximum cluster yield and most efficient fragmentation.

The initial parent's intensity is measured from TOF spectra obtained without beam crossing. In order to measure the fragments' intensity that is due exclusively to the two beam interaction, it is necessary to record TOF spectra with and without beam crossing, while keeping the same background pressure in order to account for possible metastable and/or background induced fragmentation. To obtain this, the triggering of the secondary beam nozzle is delayed by ~ 1500 μs so that no beam overlapping occurs, while the background in the scattering chamber remains unaffected. A computer program controls the digital oscilloscope and starts data recording and acquisition for, typically 2 laser shots with beam overlapping and 2 laser shots without beam overlapping. This alternating technique is repeated 200 times and was chosen in order to minimize intensity variation during the data recording and processing. With this procedure the time needed to obtain the necessary data *simultaneously for all clusters* for a given cluster series is a few minutes in comparison to a few hours required for the convectional CID with mass selection.

## 3. Results and Discussion

As mentioned above, the procedure to obtain fragmentation cross sections by using the "conventional" CID in the beam-cell arrangement is straightforward and therefore, only the results for the case of mass-selected $FeO_{10}^{+}$ cluster are given here as a representative example.

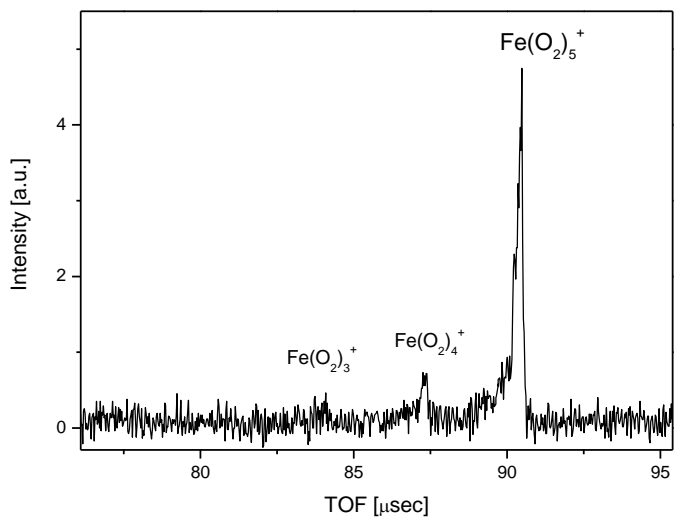

Fig. 2. Fragmentation TOF spectrum of the mass selected Fe(O$_2$)$_5^+$ cluster after collision with Kr atoms at a laboratory collision energy of 1.5 keV. The spectrum is recorded with the reflectron configuration of Fig. 1a.

In Fig. 2 is displayed the fragmentation spectrum of FeO$_{10}^+$ ion after collisions with Kr atoms at a center of mass collision energy $E_{CM}$=420 eV. The fragments, separated by the reflectron, indicate the loss of one or more oxygen molecules. The loss of molecular oxygen is observed also in all the other clusters of the homologous Fe(O$_2$)$_n^+$ (n=1-6) series investigated here. This behavior is similar to the results reported by Reilly et al.[4] for FeO$_y^+$ (y=1, 10) clusters and by Li et al.[8] for FeO$_2^+$ and FeO$_4^+$ where the loss of O$_2$ is the predominant dissociation pathway. The fragmentation cross sections defined with Eq. (7) and obtained employing the procedure described in Sec. 2.2.1 will be presented later in this section.

The mass spectra of iron oxide clusters obtained in the crossed molecular beams arrangement are presented in Fig. 3. In this case, Ne gas was used as collision partner. To investigate the outcome of the collisional activation, the TOF-mass spectra are recorded with and without beam crossing for three different retarding voltage settings: $V_{grid}= 0$ V (Fig. 3a), $V_{grid}= 1400$ V $\lesssim V_{acc}=1500$ V (Fig. 3b) and $V_{grid}= 2500$ V $>> V_{acc}$.

The mass spectra recorded in the absence of collisions and without retarding potential ($V_{grid}= 0$ V, solid curve in Fig. 3a), correspond to the species formed by mixing the laser-produced Fe-plasma plume with the $O_2$ jet. By detecting directly the charged species, the intensity of the signal is an unambiguous indication for the formation efficiency and stability of the complexes, without the drawbacks specific to post-ionization (ionization potential variation, fragmentation etc.).The spectrum in Fig. 3a (solid line) is almost identical to the one published recently[22] by our group, where a detailed presentation of iron oxide clusters distribution obtained from our source is given.

The dominant species are the oxygen rich clusters containing one Fe atom with general formula $FeO_y^+$ ($y\leq14$). The most intense peaks correspond to clusters with an even number of oxygen atoms, while ions with an odd number of oxygen atoms have low intensity, comparable or even smaller than $FeO_yH_2O^+$ complexes formed with the water impurities in the $O_2$ line.[22] Another characteristic of the clusters produced is that the intense $Fe(O_2)_n^+$ series ends-up at n=5, followed by a sudden decrease in the intensity for larger clusters. These special features of the mass spectrum have been explained previously[22] by theoretical calculations on the binding and thermodynamical properties of these clusters.

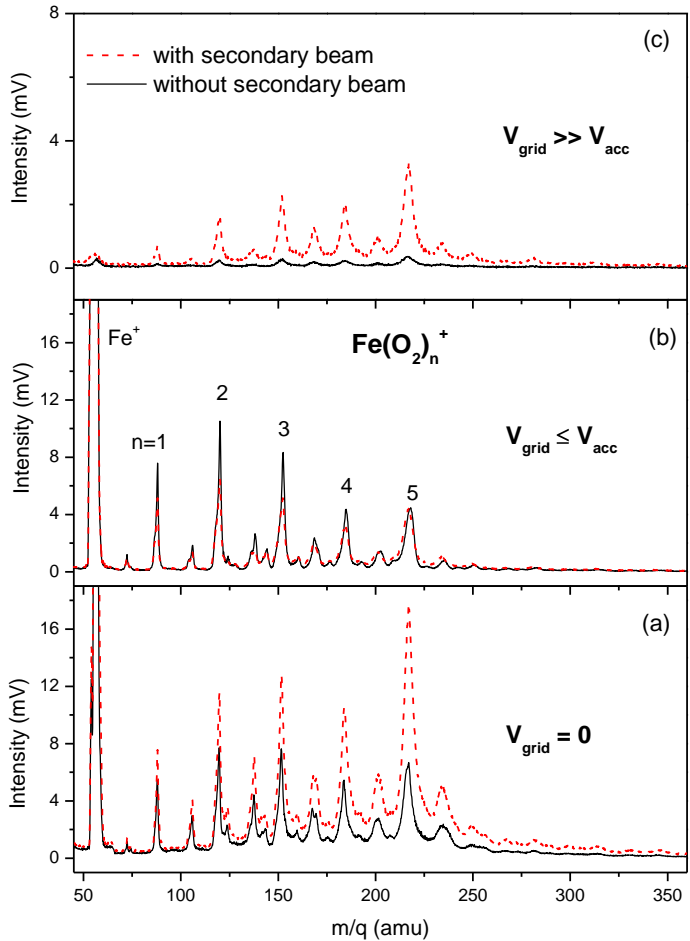

Figure 3. Mass spectra of $Fe(O_2)_n^+$ clusters (accelerated with a potential $V_{acc}$=1.5 kV) before (solid curves) and after (dotted curves) interaction with a secondary Ne-beam. Ions with kinetic energies lower than the acceleration one are rejected by a potential ($V_{grid}$) applied to a grid placed before the detector. (a) $V_{grid}$=0, (b) $V_{grid} \lesssim V_{acc}$, (c) $V_{grid} \gg V_{acc}$ (see text for details).

The collisional activation of the $Fe(O_2)_n^+$ clusters is now investigated. The dotted line in Fig. 3a represents the mass spectrum obtained for $V_{grid}$= 0 V and with Ne-secondary beam ON. Under these experimental conditions, the striking feature is an important increase of the signal intensity corresponding to all the clusters. This increase can be attributed to the emergence of new ions or neutral molecules formed by processes triggered by the collisional activation of the precursor ion as is outlined in Eqs (2-4a).

These additional particles produced are further analyzed in Fig. 3b and 3c where mass spectra are recorded after rejecting all ionic fragments ($V_{grid} \lesssim V_{acc}$, Fig. 3b) and after all ions rejection ($V_{grid} >> V_{acc}$, Fig. 3c) conditions. In Fig. 3b, where $V_{grid} \lesssim V_{acc}$, the signal corresponding to $Fe(O_2)_n^+$ clusters with secondary beam is lower than that without the secondary beam. This is due to the rejection of ionic fragments and/or multionized species (Eq.'s 2-4a). In Fig. 3c, where $V_{grid} >> V_{acc}$ only the neutral fragments are detected and constitute a significant contribution to the total signal. Small amounts of neutral particles are also produced by CID with the chamber background and/or by unimolecular (metastable) decay of hot clusters as the solid curve (without secondary beam) in Fig. 3c shows.

Using all this information and the procedure described in Sec. 2.2.2, from the mass spectra of Fig.3b and Fig.3c we can determine with Eq. (8) the individual fragmentation cross sections of $Fe(O_2)_n^+$ clusters in collisions with Ne atoms, without the need of mass selection. In Fig. 4 we plot (circles) the cross section as a function of the cluster size n for n=1, 6. Additionally, in Fig. 4 we display (squares) the fragmentation cross sections for the same cluster series obtained with the conventional CID method (Fig. 1) in a collision chamber filled with Krypton. The values originate from the evaluation described in sec. 2.2.1 [Eq. (7)] and from reflectron TOF spectra similar to the one displayed in Fig. 2. The indicative error bars shown in Fig. 4 represent the reproducibility of the data obtained at different runs, over several days.

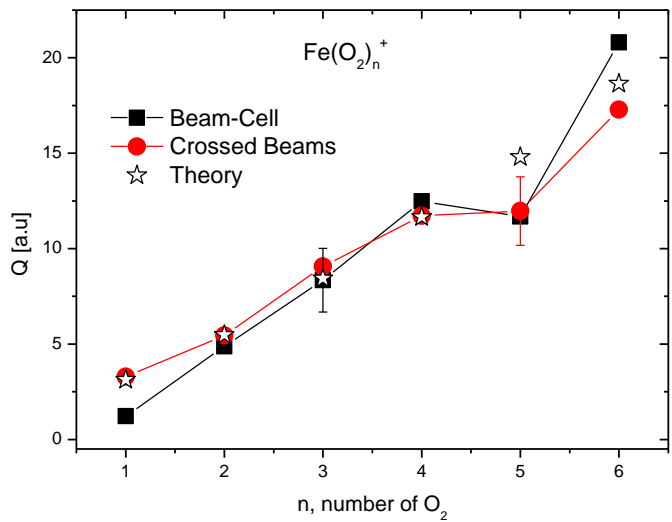

Fig. 4. The fragmentation cross section of Fe($O_2$)$_n^+$ clusters as a function of cluster size n. The results are obtained with the two different experimental techniques as described in the text. The theoretical calculations, based on the projection model from ab initio structures, are also displayed.

For comparison reasons, we scale these two curves to each other with linear least square fitting. Despite the different collision partner, both curves show the same behavior, namely an increase of the cross section with increasing cluster size, and a local minimum at n=5. It is important to notice that, while in the conventional CID technique only ionic fragments are measured, in the new approach presented here with the retarding energy analyzer all possible products from Eqs (2)-(4a) are measured. Thus, cross sections obtained with the crossed beam technique presented here are comparable to or exceed those measured with the conventional CID technique.

The very good agreement between the results obtained with the two different experimental approaches can be explained by the fact that, in all cases the center-of-mass collision energy

$$E_{CM} = \frac{m_X}{m_X + m_P} E_{LAB} \qquad (9),$$

where $m_P$ and $m_X$ are the masses of the parent and neutral collision partner respectively, is quite high for all the clusters ranging from 280 eV for $FeO_2^+$ to 110 eV for $FeO_{12}^+$. Under such highly energetic conditions (impulsive collisions), significant amount of kinetic energy is transferred to the internal energy of the cluster so that each collision leads to fragmentation. Therefore, the fragmentation cross section comes close to the integral collision cross section, which can be approximated with a hard sphere cross section and in this case quite simple theoretical approaches are possible. One such approximation is the projection method[25] that equates the integral collision cross section to the orientationally averaged cluster geometrical projection. To this end, we modified for the present case a computer program from Jarold's group[35] and calculated the averaged geometrical cross section $Q_{theo}$ of a given cluster structure, by taking into account hard sphere radii of the involved atoms of the cluster and the collision gas. As ground state equilibrium structures for the $Fe(O_2)_n^+$ clusters we consider those previously calculated by *ab initio* methods.[22] For the hard sphere radii of the atoms involved in the collision we used a simple coulombic interaction model that is represented by the so-called universal repulsive potential[36,37] given (in atomic units) by the expression:

$$V(R) = \frac{Z_1 Z_2}{R} \varphi(R/a),$$

with $a = 0.88534/(Z_1^{0.23} + Z_2^{0.23})$

The first term of the potential is the coulombic repulsion of the two nuclei and the second one is a screening function of the form:[37]

$$\varphi(x) = 0.1818\exp(-3.2x) + 0.5099\exp(-0.9423x)$$
$$+ 0.2802\exp(-0.4029x) + 0.02817\exp(-0.2016x).$$

We use this repulsive potential as the pair interaction between the target atom (Ne) and each atom (Fe and O) in the cluster. By taking into account the collision energy $E_{CM}$ for each particular cluster we calculate the distance of closest approach $R_C$ as the root of the equation $V(R_C) = E_{CM}$. For a given cluster structure taken from our previous theoretical calculations,[22] each atom in the cluster is considered as a hard sphere with fixed radius $R_C$, and the geometrical projection is calculated.

The cross sections obtained are displayed (stars) also in Fig. 4. For the sake of comparison, the three data sets are normalized to each other, despite the fact that they correspond to different noble gases as collision partners. The theoretical results reproduce quite well the evolution of the measured cross section as a function of the cluster size for all measured clusters, except for $Fe(O_2)_5$ where a relative difference of ~25% between calculation and experiment appears. This deviation can be due to the two successive approximations used in this approach. Either the first approximation, which equates the fragmentation cross section with the total collision cross section, is not entirely valid, or the second simple projection approximation applied for the calculation of cross sections fails in some cases to incorporate all the details from the cluster-atom encounter. Such effects can be the mutual shadowing or multiple collisions of the noble gas atom with different atoms of the cluster.[35,38] To account for these mechanisms, more elaborate calculation methods, such as exact hard sphere scattering, classical trajectories or scattering on electronic density isosurfaces can be employed (see Ref. 39 and references therein). These calculations - necessary for reliable structure determinations- are extremely computationally intensive and require data for interactions potentials not available in the present case and this is beyond the scope of the present investigations. Nevertheless, in the framework of the simple approximations employed here, there is a good general agreement between experiment and theory, resulting from the use of cluster structures obtained previously.[22]

Armentrout and co-workers examined the CID of mass selected Fe(L)$_n^+$ clusters, L= $O_2$[8] and N$_2$ [40] with Xe. Besides thermochemical values for these systems, the collision energy dependence of the absolute fragmentation cross sections has been investigated. From their results, we took the sum of the partial fragmentation cross sections of all channels observed at the maximum kinetic energy (4 - 15 eV) and we plot these values in Fig. 5. Additionally, we show our results from crossed beam experiments for Fe(O$_2$)$_n^+$ in collision with Ne. We normalize our values to those for L=(O$_2$)$_n$ (n=1,2) from Ref 8.

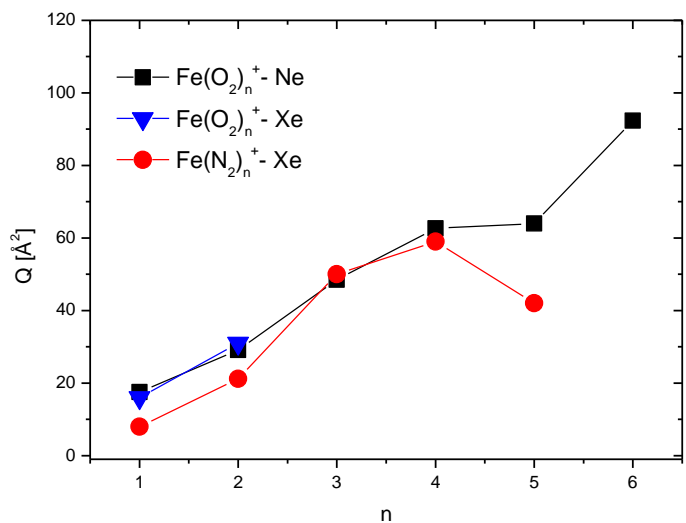

Fig. 5. The measured fragmentation cross section of the Fe(L)$_n^+$ clusters in collision with noble gases. Our data [Fe(O$_2$)$_n^+$ -Ne] are normalized to the Fe(O$_2$)$_n^+$ - Xe ones obtained in Ref. 8.

The increase in the cross section of clusters containing up to 4 molecules and then the decrease at a lower value for n=5 is similar for O$_2$ (our data) and N$_2$[40] ligands. The increase again at n=6 in our case for Fe(O$_2$)$_n^+$ cannot be compared to the other systems because of the lack of data. A lower value of the fragmentation cross section for a given cluster can be attributed to the higher stability of this in comparison to the neighboring ones.

Our experimental results are in good agreement with the results of Armentrout's group and this suggests that the new CID technique introduced here is reliable for obtaining fragmentation cross sections. In our case, where the collision energy is significantly higher than the thermal energy, the fragmentation cross section can be considered as a good approximation of collision cross sections - assumption verified by the agreement between the measured values and the geometrical cross sections obtained by theoretical calculations.

## 4. Summary and conclusion

The stability and the structure of $Fe(O_2)_n^+$ (n=1,6) clusters are investigated through CID experiments. The fragmentation cross sections after collision with noble gases are measured with two different techniques giving, practically the same results: an increase in the cross section as a function of the cluster size for n=1 to 6 with a local minimum at n=5. The cluster with n=5 was proven to have a special stability also by its higher intensity in the mass spectra. Moreover, under certain approximations, the collision cross section variation is directly representative for the variation of geometrical cross section of the molecular ions. This has been supported by the correspondence of the experimentally measured cross section with the values obtained by geometrical projection of theoretical structures, obtained by *ab initio* calculations in the case of $Fe(O_2)_n^+$ adducts. Finally, a new approach based on crossed molecular beams for CID studies, without the need of mass selection, has been presented and confirmed by comparison with an established CID method, with data available in the literature and with results obtained by theoretical calculations.

**Acknowledgment**

Support from a grand of Romanian National Authority for Scientific Research, CNCS-UEFISCDI, project number PN-II-RU-TE-2011-3-0174, is gratefully acknowledged.

**References**

(1) Cox, D. M.; Trevor, D. J.; Whetten, R. L.; Rohlfing, E. A.; Kaldor, A. Magnetic Behavior of Free-iron and Iron Oxide Clusters. *Phys. Rev. B* **1985,** *32*, 7290-7298.

(2) Wang, L.-S.; Wu, H.; Desai, S. R. Sequential Oxygen Atom Chemisorption on Surfaces of Small Iron Clusters. *Phys. Rev. Lett.* **1996**, *76*, 4853-4856.

(3) Wang, Q.; Sun, Q.; Sakurai, M.; Yu, J. Z.; Gu, B. L.; Sumiyama, K.; Kawazoe, Y. Geometry and Electronic Structure of Magic Iron Oxide Clusters. *Phys. Rev. B* **1999**, *59*, 12672-12677.

(4) Reilly, N. M.; Reveles, J. U.; Johnson, G. E.; del Campo, J. M.; Khanna, S. N.; Köster, A. M.; Castleman Jr., A. W. Experimental and Theoretical Study of the Structure and Reactivity of $Fe_mO_n^+$ (m = 1, 2; n = 1 - 5) with CO. *J. Phys. Chem. C* **2007**, *111*, 19086-19097.

(5) Molek, K. S.; Anfuso-Cleary, C.; Duncan, M. A. Photodissociation of Iron Oxide Cluster Cations. *J. Phys. Chem. A* **2008**, *112*, 9238-9247.

(6) Xie, Y.; Dong, F.; Heinbuch, S.; Rocca, J. J.; Bernstein, E. R. Investigation of the Reactions of Small Neutral Iron Oxide Clusters with Methanol. *J. Chem. Phys.* **2009**, *130*, 114306 - 114306-11.

(7) Yin, S.; Xue, W.; Ding, X.-L.; Wang, W.-G.; He, S.-G.; Ge, M.-F. Formation, Distribution, and Structures of Oxygen-Rich Iron and Cobalt Oxide Clusters. *Int. J. Mass Spectrom.* **2009**, *281*, 72-78.

(8) Li, M.; Liu, S.-R.; Armentrout, P. B. Collision-Induced Dissociation Studies of $Fe_mO_n^+$: Bond Energies in Small Iron Oxide Cluster Cations, $Fe_mO_n^+$ (m = 1-3, n = 1-6). *J. Chem. Phys.* **2009**, *131*, 144310 - 144310-16.

(9) Xue, W.; Yin, S.; Ding, X.-L.; He, S.-G.; Ge, M.-F. Ground State Structures of $Fe_2O_{4-6}^+$ Clusters Probed by Reactions with $N_2$. *J. Phys. Chem. A* **2009**, *113*, 5302-5309.

(10) Shin, D. N.; Matsuda, Y.; Bernstein, E. R. On the Iron Oxide Neutral Cluster Distribution in the Gas Phase. I. Detection through 193 nm Multiphoton Ionization. *J. Chem. Phys.* **2004**, *120*, 4150-4156.

(11) Shin, D. N.; Matsuda, Y.; Bernstein, E. R. On the Iron Oxide Neutral Cluster Distribution in the Gas Phase. II. Detection through 118 nm Single Photon Ionization, *J. Chem. Phys.* **2004**, *120*, 4157-4164

(12) Riley, S. J.; Parks, E. K.; Nieman, G. C.; Pobo, L. G.; Wexler, S. Metal-Deficient Iron Oxide Clusters Formed in the Gas Phase. *J. Chem. Phys.* **1984**, *80*, 1360-1362.

(13) Schröder, D.; Jackson, P.; Schwarz, H. Dissociation Patterns of Small $Fe_mO_n^+$ (m = 1–4, n ≤ 6) Cluster Cations Formed Upon Chemical Ionization of $Fe(CO)_5/O_2$ Mixtures. *Eur. J. Inorg. Chem.* **2000**, *2000*, 1171-1175.

(14) Reilly, N. M.; Reveles, J. U.; Johnson, G. E.; Khanna, S. N.; Castleman Jr., A. W. Experimental and Theoretical Study of the Structure and Reactivity of $Fe_{1-2}O_{\leq 6}^-$ Clusters with CO. *J. Phys. Chem. A* **2007**, *111*, 4158-4166.

(15) Shiroishi, H.; Oda, T.; Hamada, I.; Fujima, N. Structure and Magnetism on Iron Oxide Clusters $Fe_nO_m$ (n=1-5): Calculation from First Principles. *Eur. Phys. J. D* **2003**, *24*, 85-88.

(16) Atanasov, M. Theoretical Studies on the Higher Oxidation States of Iron. *Inorg. Chem.* **1999**, *38*, 4942-4948.

(17) Gutsev, G. L.; Khanna, S. N.; Rao, B. K.; Jena, P. Electronic Structure and Properties of $FeO_n$ and $FeO_n^-$ Clusters. *J. Phys. Chem. A* **1999**, 103, 5812- 5822.

(18) López, S.; Romero, A. H.; Mejía-López, J.; Mazo-Zuluaga, J.; Restrepo, J. Structure and Electronic Properties of Iron Oxide Clusters: A First-Principles Study. *Phys. Rev. B* **2009**, *80*, 085107 - 085107-10.

(19) Gutsev, G. L.; Weatherford, C. A.; Pradhan, K.; Jena, P. Structure and Spectroscopic Properties of Iron Oxides with the High Content of Oxygen: $FeO_n$ and $FeO_n^-$ (n=5-12). *J. Phys. Chem. A* **2010**, *114*, 9014-9021.

(20) Palotás, K.; Andriotis, A. N.; Lappas, A. Structural, Electronic, and Magnetic Properties of Nanometer-Sized Iron-Oxide Atomic Clusters: Comparison between GGA and GGA+U Approaches. *Phys. Rev. B* **2010**, 81, 075403 - 075403-14.

(21) Reilly, N. M.; Reveles, J. U.; Johnson, G. E.; Khanna, S. N.; Castleman Jr., A. W. Influence of Charge State on the Reaction of $FeO_3^{+/-}$ with Carbon Monoxide. *Chem. Phys. Lett.* **2007**, *435*, 295-300.

(22) Mpourmpakis, G.; Velegrakis, M.; Mihesan, C.; Andriotis, A. N. Symmetry-Switching Molecular Fe(O$_2$)$_n^+$ Clusters. *J. Phys. Chem. A* **2011**, *115*, 7456-7460.

(23) Begemann, W.; Dreihöfer, S.; Meiwes-Broer, K. H.; Lutz, H. O. Sputtered Metal Cluster Ions: Unimolecular Decomposition and Collision Induced Fragmentation. *Z. Phys. D: Atoms, Molecules and Clusters* **1986**, *3*, 183-188.

(24) Deng, H. T.; Kerns, K. P.; Castleman, Jr., A. W. Formation, Structures, and Reactivities of Niobium Oxide Cluster Ions. *J. Phys. Chem.* **1996**, *100*, 13386 -13392.

(25) von Helden, G.; Hsu, M. T.; Gotts, N.; Bowers M. T. Carbon Cluster Cations with up to 84 Atoms: Structures, Formation Mechanism, and Reactivity. *J. Phys. Chem.* **1993**, *97*, 8182- 8192.

(26) Jarrold, M. F. Drift Tube Studies of Atomic Clusters. *J. Phys. Chem.* **1995**, *99*, 11-21.

(27) Velegrakis, M.; Sfounis, A. Formation and Photodecomposition of Cationic Titanium Oxide Clusters. *Appl. Phys. A,* **2009**, *97*, 765-770.

(28) Jadraque, M.; Sierra, B.; Sfounis, A.; Velegrakis, M. Photofragmentation of Mass-Selected Titanium Oxide Cluster Cations. *Appl. Phys. B* **2010**, *100*, 587-590.

(29) Velegrakis, M.; Massaouti, M.; Jadraque, M. Collision-Induced Dissociation Studies on Gas-Phase Titanium Oxide Cluster. *Appl. Phys. A* **2012**, 108, 127-131.

(30) Lüder, C.; Velegrakis, M. Photofragmentation Spectrum of the Sr$^+$Ar Complex. *J. Chem. Phys.* **1996**, *105*, 2167-2176

(31) Lüder, C.; Georgiou, E.; Velegrakis, M. Studies on the Production and Stability of Large C$_N^+$ and M$_x^+$R$_N$ (M = C, Si, Ge and R = Ar, Kr) Clusters. *Int. J. Mass Spectrom. Ion Processes* **1996**, *153*, 129-138.

(32) Witkowicz, E.; Linnartz, H.; de Lange, C.A.; Ubachs, W.; Sfounis, A.; Massaouti, M.; Velegrakis, M. Mass Spectrometric and Laser Spectroscopic Characterization of a Supersonic Planar Plasma Expansion. *Int. J. Mass Spectrom.* **2004**, *232*, 25-29.

(33) J.J.H. van den Biesen. Elastic scattering I: Integral cross sections. In G. Scoles, editor, Atomic and Molecular Beam Methods. Oxford University Press, New York, NY, 1988.

(34) Barat, M.; Brenot, J. C.; Fayeton, J. A.; Picard, Y. J. Absolute Detection Efficiency of a Microchannel Plate Detector for Neutral Atoms. *Rev. Sci. Instrum.* **2000**, *71*, 2050-2052.

(35) Shvartsburg A.; Jarrold, M. F. An Exact Hard-Spheres Scattering Model for the Mobilities of Polyatomic Ions. *Chem. Phys. Lett.* **1996**, *261*, 86-91; http://www.indiana.edu/~nano/software.html

(36) O'Connor, D.J.; Biersack, J.P. Comparison of Theoretical and Empirical Interatomic Potentials. *Nucl. Instrum. Methods Phys. Res., Sect. B* 1986, 15, 14-19.

(37) Biersack, P.; Ziegler, J.F. The Stopping and Range of Ions in Solids. Springer Ser. Electrophysics, Berlin, 1982, 10, 122-156.

(38) Weis, P.; Welz,O.; Vollmer, E.; Kappes, M. M. Structures of Mixed Gold-Silver Cluster Cations ($Ag_mAu_n^+$, m+n<6): Ion Mobility Measurements and Density-Functional Calculations. *J. Chem. Phys.* **2004**, *120*, 677-684.

(39) Shvartsburg A. ; Liu, B.; Jarrold, M. F.; Ho, K-M. Modeling Ionic Mobilities by Scattering on Electronic Density Isosurfaces: Application to Silicon Cluster Anions. *J. Chem. Phys*. **2000**, *112*, 4517-4526.

(40) Tjelta B. L; Armentrout, P.B. Gas-Phase Metal Ion Ligation: Collision-Induced Dissociation of $Fe(N_2)_x^+$ (x = 1-5) and $Fe(CH_2O)_x^+$ (x = 1-4). *J. Phys. Chem. A* **1997**, *101*, 2064-2073.

**TOC Image**

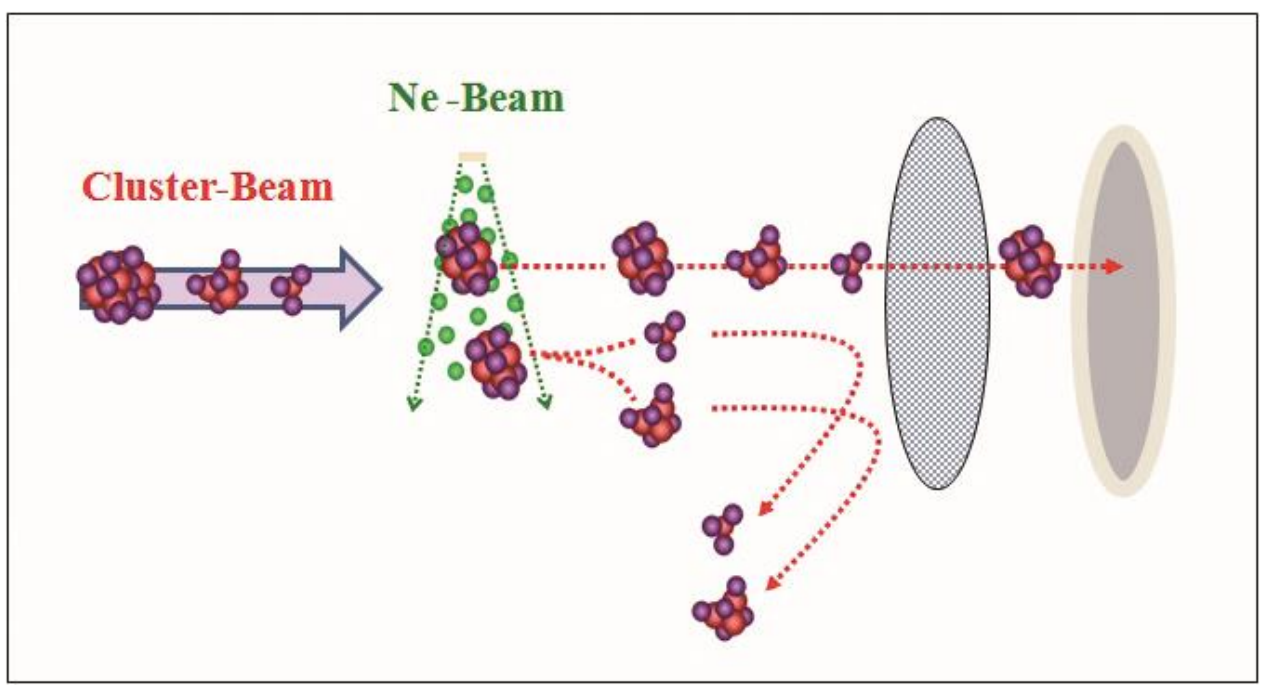